\documentclass[10pt]{iopart}
\usepackage{iopams}

\expandafter\let\csname equation*\endcsname=\relax
\expandafter\let\csname endequation*\endcsname=\relax
\usepackage{amsmath}

\usepackage{amssymb}
\usepackage{amsfonts,latexsym}
\usepackage{bm}
\usepackage[mathcal]{euscript}
\usepackage{graphicx}
\usepackage{epsfig}
\usepackage{color}
\usepackage{hyperref}

\newcommand{\be}{\begin{equation}}
\newcommand{\ee}{\end{equation}}
\newcommand{\bea}{\begin{eqnarray}}
\newcommand{\eea}{\end{eqnarray}}

\newcommand{\la}{\langle}
\newcommand{\ra}{\rangle}

\begin{document}

\title[Collective oscillations in a 3D spin model]{Collective oscillations in a three-dimensional spin model with non-reciprocal interactions}

\date{\today}
\author{Laura Guislain and Eric Bertin}

\address{Univ.~Grenoble Alpes, CNRS, LIPhy, F-38000 Grenoble, France}


\begin{abstract}
We study the onset of collective oscillations at low temperature in a three-dimensional spin model with non-reciprocal short-range interactions.
Performing numerical simulations of the model, the presence of a continuous phase transition to global oscillations is confirmed by a finite-size scaling analysis.
By systematically varying the interaction range, we show that collective oscillations in this spin model actually result from two successive phase transitions: a mean-field phase transition over finite-size neighborhoods, which leads to the emergence of local noisy oscillators, and a synchronization transition of local noisy oscillators, which generates coherent macroscopic oscillations. 
Using a Fokker-Planck equation under a local mean-field approximation, we derive from the spin dynamics coupled Langevin equations for the complex amplitudes describing noisy oscillations
on a mesoscopic scale. The phase diagram of these coupled equations is qualitatively obtained from a fully-connected (mean-field) approximation.
This analytical approach allows us to clearly disentangle the onset of local and global oscillations, and to identify the two main control parameters, expressed as combinations of the microscopic parameters of the spin dynamics, that control the phase diagram of the model.
\end{abstract}

\section{Introduction}

The onset of collective oscillations in far-from-equilibrium systems is a ubiquitous phenomenon
\cite{nicolis_dissipative_1986}. A first idea to model the emergence of collective oscillations is to consider a large number of coupled oscillators, which reach a synchronized state when interactions are strong enough. This is the basic mechanism at play in the Kuramoto model \cite{acebron_kuramoto_2005,gupta_kuramoto_2014}, which considers oscillators with distributed frequencies. However, not all systems exhibiting collective oscillations explicitly consist of a collection of coupled microscopic oscillators. In many cases, individual degrees of freedom do not oscillate in the absence of interactions. This is the case for different types of experimental systems like biochemical clocks \cite{Cao_free_energy2015,nguyen_phase_2018,Aufinger_complex2022} or droplets in a fluid binary mixture \cite{devailly_phase_2015}, as well as in stochastic models lacking detailed balance like kinetic spin models \cite{collet_rhythmic_2016,de_martino_oscillations_2019,daipra_oscillatory_2020,Sinelschikov_emergence_2023},
socio-economic models \cite{Gualdi2015,yi_symmetry_2015}, or models of population dynamics \cite{andrae_entropy_2010,Duan_Hopf2019}.

For infinitely large systems, for which fluctuations can be neglected, the onset of spontaneous oscillations can be accounted for in the frame of dynamical system theory, and is described by 
a Hopf bifurcation \cite{crawford_introduction_1991}.
In contrast, oscillations occurring in mesoscopic systems are intrinsically noisy 
(see, e.g., \cite{Fei_design2018} in the context of biochemical clocks), notably leading to a loss
of phase coherence over a finite time \cite{gaspard_correlation_2002,barato_cost_2016,barato_coherence_2017,oberreiter_universal_2022,remlein_coherence_2022}.
Such noisy oscillations may be phenomenologically described as a stochastic Hopf bifurcation \cite{Sagues2007,Xu_Langevin2020}.
A description rooted in the microscopic dynamics and based on a large deviation approach has been proposed recently
\cite{guislain_nonequil2023,guislain2024discontinuous} in order to generalize the usual Landau theory of phase transitions to the case of the onset of spontaneous oscillations.
This Landau approach of the phase transition complements earlier approaches based on nonequilibrium thermodynamic concepts, like the presence of a singularity in the entropy production
\cite{crochik_entropy_2005,xiao_entropy_2008,xiao_stochastic_2009,barato_entropy_2012,tome_entropy_2021,noa_entropy_2019,martynec_entropy_2020,seara_irreversibility_2021}.

However, the Landau approach is restricted to mean-field systems, and a natural question would be to try to characterize the onset of spontaneous oscillations in finite-dimensional systems, as recently reported for instance in experiments on cell monolayers \cite{petrolli2019,peyret2019}, in populations of biological cells \cite{Kamino_fold2017,Wang_emergence2019},
or in systems of active particles subjected to non-reciprocal interactions \cite{saha_scalar_2020,you_nonreciprocity_2020,martin2024transition}.
This nonequilibrium phase transition has been characterized using renormalization group methods
in models of identical coupled noisy oscillators
\cite{risler.etal.2004,risler.etal.2005}, confirming in particular the existence of a continuous phase transition in three dimensions.
In two dimensions, the presence of defects has been recently shown to destroy the global oscillating phase \cite{avni.etal.2023}.

In this paper, we analyze a three-dimensional spin model with non-reciprocal interactions exhibiting collective oscillations. We show using both numerical and analytical methods that effective local oscillators emerge on a mesoscopic scale from the microscopic dynamics through a finite-size, mean-field phase transition. In a second stage, these local noisy oscillators may synchronize on a global scale, leading to the onset of collective oscillations. We describe in details the interplay of these two types of phase transitions. An analytical approach allows us in particular to identify the few key microscopic parameters that control the macroscopic phase diagram of the model.
The article is organized as follows.
In Sec.~\ref{sec:model}, the model is introduced and the phase transition to a globally oscillating state is characterized numerically.
The link with the synchronization transition of coupled oscillators is explored in Sec.~\ref{sec:link:synchro}, where the mean-field phase diagram of coupled noisy oscillators is also characterized. Finally, the influence of the microscopic parameters of the three-dimensional spin models on the global phase diagram obtained from coupled noisy oscillators is studied in Sec.~\ref{sec:comparison}.

\section{Collective oscillations in a spin model with short-range interactions}
\label{sec:model}

\subsection{Model and key observables}

\subsubsection{Definition of the model}

\begin{figure}
    \centering
    \includegraphics[width=1\linewidth]{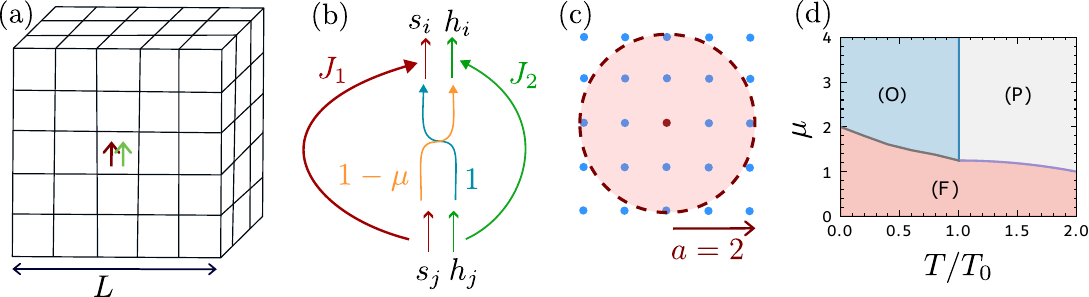}
    \caption{(a) Schematic of the 3D model: $N$ spins and fields are placed on a cubic lattice of size $L$. (b) Diagram of interactions between spins and fields. (c) Representation for $a=2$ (2D section of the 3D lattice) of interaction range, corresponding to a spherical neighborhood of radius $a$ around each site is considered in the interactions. 
    (d) Phase diagram for infinite range interactions ($n(a)=N\to\infty$) for $J_1=0$ and $J_2=1$, from \cite{guislain_nonequil2023,guislain2024discontinuous}; (P) indicates the paramagnetic phase, (F) the ferromagnetic phase and (O) the oscillating phase.}
    \label{fig:MF:phase:diagram}
\end{figure}

We consider a model with $N=L^d$ spins $s_i$ and fields $h_i$ in dimension $d=3$ placed on a cubic lattice, as illustrated in Fig.~\ref{fig:MF:phase:diagram}(a).  Spins and fields interact with neighboring sites. We consider spherical neighborhoods of radius $a$, where $a$ can be non-integer [see Fig.~\ref{fig:MF:phase:diagram}(c)].
We denote $n(a)$ the number of neighbors, and $d(i,j)$ the distance between two sites $i$ and $j$.
We consider a single spin $s_i$ or field $h_i$ flip dynamic. The transition rates $W_s(i)$ and $W_h(i)$ to flip a spin $s_i=\pm1$ or a field $h_i=\pm1$ are given by 
\bea
W_s(i) &=& [1+\exp(\Delta E_s(i)/T)]^{-1},\\
W_h(i) &=& [1+\exp( \Delta E_h(i)/T)]^{-1},
\eea
where $T$ is the bath temperature; $\Delta E_s(i)$ and $\Delta E_h(i)$ are the variations of interaction energies $E_s$ and $E_h$ when flipping spin $s_i$ or field $h_i$ respectively.
These interaction energies are defined as [see Fig.~\ref{fig:MF:phase:diagram}(b)]:
\bea
E_s &=& -\frac{1}{n(a)}\sum_{d(i, j)\leq a}\left(\frac{J_1}{2}s_is_j +s_ih_j+\frac{J_2}{2}h_ih_j\right),\\
E_h &=& E_s+ \frac{\mu}{n(a)}\sum_{d(i, j)\leq a}s_ih_j,
\eea
where the sum $\sum_{d(i,j)\leq a}$ is performed over all pairs $(i,j)$ separated by a distance $d(i,j)$ less than or equal to $a$.
The coupling constants $J_1$ and $J_2$ control the ferromagnetic interactions between the spins and between the fields. 
We take $J_1=0$ for the stochastic simulations, and we briefly comment on the effect of $J_1$ in \ref{app:J1}. 
When $\mu=0$, $E_h=E_s$ so that detailed balance is verified. Hence the parameter $\mu$ controls the deviation from equilibrium, through the non-reciprocity of interactions.

The limit where all spins and fields interact, corresponding to $n(a)=N$, is studied in \cite{guislain_nonequil2023, guislain2024discontinuous}. 
In this limit, temporal oscillations appear for $T<T_0$, with
\be \label{def:T0}
T_0=\frac{J_1+J_2}{2}
\ee
provided that $\mu>\mu_c=1+(J_1+J_2)^2/4$. At high $\mu$ and low $T$, the system is then in an oscillating phase. At a fixed $\mu$, the transition from the disordered to the oscillating phase corresponds to a Hopf bifurcation in the deterministic limit. 
The phase diagram of the model is shown in Fig.~\ref{fig:MF:phase:diagram}(d). The low values of $\mu$ and $T$ correspond to a ferromagnetic phase with two ferromagnetic points $(m_0, h_0), (-m_0, -h_0)$.

\subsubsection{Order parameters characterizing the phase transition}
The oscillating phase can be described using two order parameters. 
First, we define the magnetization as
\be m=N^{-1}\sum_{i=1}^N s_i\ee
and calculate $\langle m^2\rangle$ from stochastic simulations (see below). A non-zero value of $\langle m^2\rangle$ corresponds to an ordered (oscillating or ferromagnetic) phase, while $\langle m^2\rangle=0$ in the limit $N\to\infty$ (or, more precisely, $\langle m^2\rangle \sim N^{-1}$ for large $N$) corresponds to a paramagnetic phase. 
To distinguish a ferromagnetic phase from an oscillating phase, we use the smoothed stochastic derivative of the magnetization $\dot{m}$ introduced in \cite{guislain_nonequil2023} and defined as follows 
\be \label{def:mdot}\dot{m}(\{s_i\},\{h_i\})=-\frac{2}{N}\sum_{i=1}^N s_i W_s(i).\ee
In the limit $N\to\infty$, the mean-square stochastic derivative of the magnetization $\langle \dot{m}^2\rangle$ is non-zero only in the oscillating phase. Unlike the magnetization, the stochastic derivative of the magnetization is expensive to compute numerically, as all $W_s(i)$ must be evaluated at each step.
Thus, when extensive simulations need to be performed, like to determine the critical temperature $T_c(\mu)$, it may be more convenient to use the magnetization as order parameter, after checking from the trajectories $m(t)$ or from $\langle\dot{m}^2\rangle$ that oscillations are present.

\subsubsection{Motivation for choosing to vary the range of interaction}
\label{subsubsec:a1}
First, we briefly consider the simplest version of the model where spins and fields interact only with their nearest neighbors, corresponding to $a=1$ (and therefore $n(a)=6$ in three dimensions) in the previously introduced model. 
We have run simulations up to $L=14$ ($N=L^3=2744$) for different temperatures $T$ and different $\mu$ (simulation results are presented in \ref{app:simu:a1} on Fig.~\ref{app:fig:a1}) for $J_1=0$ and $J_2=1$. For $\mu \geq 2$ and all temperatures, we observe that $\langle m^2\rangle$ decreases in $N^{-1}$, indicating the absence of an ordered phase.
Thus, we observe that when considering nearest-neighbor interactions ($a=1$) in this model and for the choice of $J_1=0$ and $J_2=1$, the oscillating phase does not exist, whereas it does when considering infinite-range interactions. The choice of $J_1$, $J_2$ is arbitrary here, but simulations for lower $J_2$ values give similar results. 
We note, however, that the absence of oscillations when considering nearest-neighbor interactions does not seem to be a universal property of spin models with non-reciprocal interactions \cite{avni.etal.2023}, and it depends on the specific definition of the model. 

To sum up, the model exhibits collective oscillations when the interaction range is infinite, but the oscillations disappear at very short range. We now vary the range of interaction $a$ in order to understand how collective oscillations appear. Intuitively, if the range of interaction is large enough, noisy oscillators are expected locally, and then the onset of system-wide oscillations
is likely to be similar to a synchronization transition of noisy oscillators.

\subsection{Numerical results for $a>1$}
\subsubsection{Presence of collective oscillations}
\begin{figure}[t]
    \centering
    \includegraphics{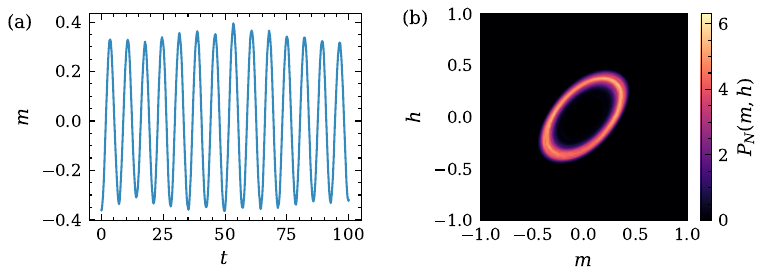}
    \caption{Evidence of collective oscillations. (a) Trajectory $m(t)$ and (b) probability density $P_N(m, h)$ obtained from numerical simulations of the three-dimensional spin model. Parameters: $J_1=0$, $J_2=1$, $\mu=2$, $T=0.2$, $a=2$ and $L=21$.}
    \label{fig:traj_hist}
\end{figure}
\begin{figure}[t]
    \centering
    \includegraphics{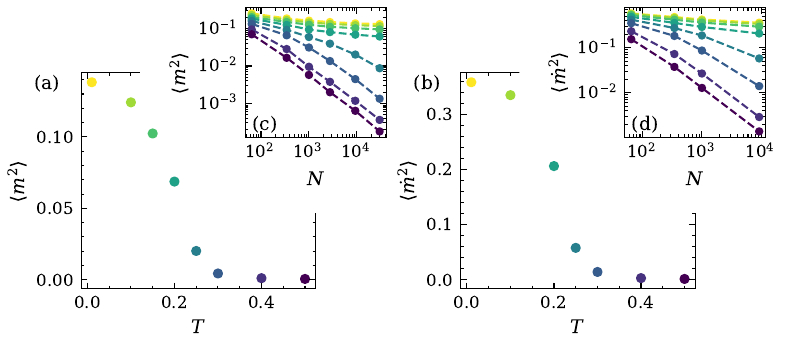}
    \caption{Evidence of a phase transition towards collective oscillations. (a) $\langle m^2\rangle$ and (b) $\langle \dot{m}^2\rangle$ for different temperatures $T$ for $L=21$. (c) $\langle m^2\rangle$ and (d) $\langle \dot{m}^2\rangle$ for different system sizes $N=L^3$; each color corresponds to different temperatures [shown in (a) and (b)]. Parameters: $J_1=0$, $J_2=1$, $\mu=2$ et $a=2$.}
    \label{fig:simu:a2:mmd}
\end{figure}


We now consider a larger interaction range $a>1$. Before studying the influence of varying the various parameters ($\mu$, $a$, $J_1$ and $J_2$), we fix $J_1=0$, $J_2=1$, $\mu=2$ and $a=2$ (corresponding to a neighborhood of $n(a)=32$ sites) and show that the model exhibits, for a certain temperature range, collective temporal oscillations. 
A trajectory $m(t)$ and the probability density $P_N(m, h)$ is shown in Fig.~\ref{fig:traj_hist} for $T=0.2$ (low temperature) and a system size $L=21$ (corresponding to $N=21^3=9261$). We observe spontaneous oscillations of the magnetization. 
Once the presence of oscillations is confirmed by a nonzero measured value of $\langle \dot{m}^2\rangle$ on a given system size, the amplitude of oscillations is precisely characterized as a function of
system size $L$ by measuring $\langle m^2\rangle$, which is less costly numerically.
In Figs.~\ref{fig:simu:a2:mmd}(a) and (b), we plot $\langle m^2\rangle$ and $\langle \dot{m}^2\rangle$ as a function of temperature for $L=21$. In order to verify that the observed oscillations are not finite size effects, we study the evolution of the two order parameters as a function of the system size $N=L^3$ for each temperature [see Figs.~\ref{fig:simu:a2:mmd}(c) and (d)].
We observe that for high temperatures, $\langle m^2\rangle\sim N^{-1}$ and $\langle \dot{m}^2\rangle\sim N^{-1}$ corresponding to a paramagnetic phase, whereas for low temperatures we have $\langle m^2\rangle\neq 0$ and $\langle \dot{m}^2\rangle\neq 0$, which corresponds to a phase with temporal oscillations of the magnetization.

\begin{figure}[t]
    \centering
    \centering    \includegraphics{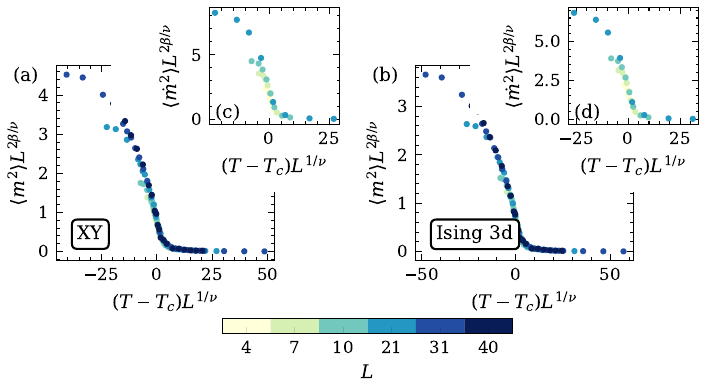}
    \caption{(a), (b) Finite-size scaling analysis of the mean-squared magnetization $\la m^2\ra$: (a) with the exponents of the 3D XY-model and (b) with the exponents of the 3D Ising model.
    (c), (d) Finite-size scaling analysis of the mean-squared derivative of the magnetization $\la \dot{m}^2\ra$, with the 3D XY-model (c) or 3D Ising (d) exponents.
    Parameters: $J_1=0$, $J_2=1$, $\mu=2$ et $a=2$.}
    \label{fig:simu:a2:mmd:scal}
\end{figure}
To confirm the existence of a continuous phase transition and try to determine its universality class, we use a finite-size scaling analysis of the magnetization \cite{privman.1990,fisher.1972}. We plot $\langle m^2\rangle L^{2\beta/\nu}$ as a function of $(T-T_c)L^{1/\nu}$ where $\beta$ and $\nu$ are the critical exponents characterizing the magnetization and the correlation length respectively. We then seek to obtain a collapse of the curves associated with different sizes $N=L^3$. 
For comparison with the 3D XY-model, we use $\beta=0.34$ and $\nu=0.66$ \cite{gottlob.hasenbusch.1993}, while for the 3D Ising model we use $\beta=0.31$ and $\nu=0.64$ \cite{huang.2008}. We present the results in Fig.~\ref{fig:simu:a2:mmd}(a), (b). 
Both sets of exponent values give a satisfactory data collapse, which confirms the existence of a continuous phase transition to an oscillating state.
We also perform a similar finite-size scaling analysis on the derivative $\dot{m}$ of the magnetization, by plotting $\langle \dot{m}^2\rangle L^{2\beta/\nu}$ as a function of $(T-T_c)L^{1/\nu}$
in Fig.~\ref{fig:simu:a2:mmd}(c), (d). 
Due to the higher numerical cost of the evaluation of $\langle \dot{m}^2\rangle$ as compared to $\langle m^2\rangle$, less data are available, but the resulting collapse is still reasonable.
Given that exponent values of the three-dimensional XY and Ising models are very close,
our finite-size scaling analysis does not allow us to distinguish between these two universality classes. 
While in some situations the physics of the problem may help in selecting the correct universality class, here both classes may be considered as physically relevant:
microscopically, the model is composed of Ising spins, but at a mesoscopic scale oscillators similar to the continuous spins of the XY model (driven at a given angular frequency) also appear, as discussed below.
Note that we find a similar behavior to previous studies of discrete oscillators \cite{avni.etal.2023, wood.etal.2006, wood.etal.2006a}, in which simulations at larger system sizes were performed,
and where it was found that the critical exponents slowly approach the XY universality class when increasing system size.

In summary, we observe that our model has no oscillatory phase when considering only nearest-neighbor interactions ($a=1$), but by increasing the interaction range $a$, an oscillatory phase appears, having critical properties similar to those of noisy interacting oscillators \cite{risler.etal.2004, risler.etal.2005} and to the oscillatory phase observed in similar discrete models \cite{avni.etal.2023, wood.etal.2006}.

\subsubsection{Influence of the range of interaction $a$}
\begin{figure}
    \centering
    \includegraphics{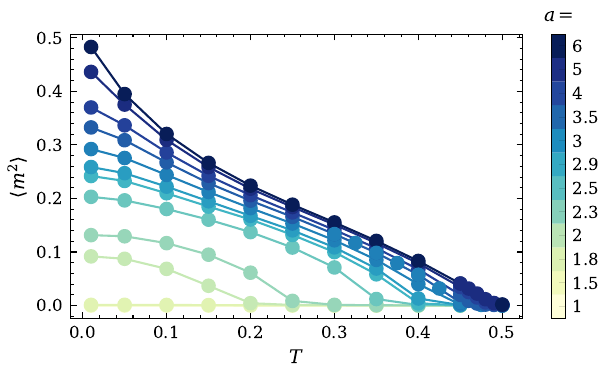}
    \caption{Influence of the interaction range $a$: $\langle m^2\rangle$ as a function of temperature $T$ for different values of $a$. Parameters: $J_1=0$, $J_2=1$, $\mu=2$, $L=31$.}
    \label{fig:Tc:a}
\end{figure}

We take $\mu=2$ and vary the interaction range $a$. 
While no oscillating phase is found for $a=1$, for $a=2$ we find a transition to a spontaneously oscillating phase at $T_c\approx0.23$. 
In Fig.~\ref{fig:Tc:a}, we plot $\langle m^2\rangle$ as a function of temperature for different values of the interaction range $a$ for $L=31$ ($N=L^3=29791$). An oscillating phase appears for $a=1.8$. The greater the interaction range, the greater the amplitude of the oscillations and the more extended the oscillating phase (i.e., the higher the critical temperature). 

\begin{figure}
    \centering
    \includegraphics{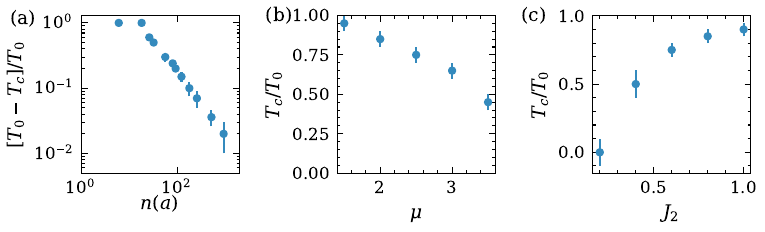}
    \caption{Evolution of the critical temperature $T_c$ with the different parameters of the model ($T_c$ is set against the critical temperature $T_0$ of the infinite range model): (a) with $n(a)$ for $\mu=2$, $J_1=0$ and $J_2=1$; (b) with $\mu$ for $a=3$, $J_1=0$ and $J_2=1$; (c) with $J_2$ for $J_1=0$, $a=3$ and $\mu=2$. Each critical temperature is obtained using the dependence with $N=L^3$ of $\langle m^2\rangle$ (for sizes up to $L=41$).
    }
    \label{fig:Tcnumres}
\end{figure}
To identify the dependence of the critical temperature, corresponding to the onset of oscillations, on the interaction range, we study the dependence of $\langle m^2\rangle$ with system size, for sizes up to $L=41$ ($N=L^3=68921$).
The critical temperature is displayed in Fig.~\ref{fig:Tcnumres}(a) as a function of the neighborhood size $n(a)$, and we find that 
\be \frac{T_0-T_c(a)}{T_0}\sim \frac{1}{n(a)},\ee
where $T_0=(J_1+J_2)/2$ is the critical temperature of the model with infinite interaction range.

\subsubsection{Influence of the non-reciprocity parameter $\mu$}
\begin{figure}
    \centering
    \includegraphics{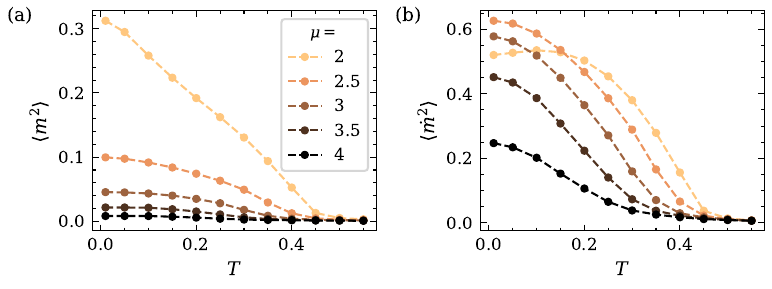}
    \caption{Influence of the non-reciprocity parameter $\mu$ on (a) $\langle m^2\rangle$ and (b) $\langle \dot{m}^2\rangle$ as a function of temperature for different values of $\mu$ indicated in the legend, for $L=14$. Same colorcode on both panels. Parameters: $J_1=1$, $J_2=1$, $a=3$. 
    }
    \label{fig:simu:mu}
\end{figure}
We now take $a=3$ and study the effect of the non-reciprocity parameter $\mu$. In Fig.~\ref{fig:simu:mu}, we plot $\langle m^2\rangle$ and $\langle \dot{m}^2\rangle$ as a function of temperature $T$ for $L=14$ and for various high values of $\mu$ ($\mu>2$). We comment on lower values of $\mu$ in \ref{app:discussion:mu:petit}.
We observe that $\langle m^2\rangle$ and $\langle \dot{m}^2\rangle$ decrease with $\mu$ for a fixed temperature and increase when $T$ decreases at a fixed $\mu$. 
In the infinite interaction range limit, $m^2\sim 1/\mu^2$ while $\dot{m}^2\sim1/\mu$, which explains why $\langle m^2\rangle$ decreases more strongly with $\mu$ than $\langle \dot{m}^2\rangle$ in the three-dimensional model.

To distinguish between finite size effects ($\langle m^2\rangle\sim N^{-1}$) and small values of $\langle m^2\rangle$ close to the transition (which remain non-zero for $N\to\infty$), we calculate $\langle m^2\rangle$ for different system sizes up to $L=31$ [see \ref{app:eval:tc:mu:varie}] and plot the resulting phase diagram on Fig.~\ref{fig:Tcnumres}(b). As $\mu$ increases, the critical temperature decreases, whereas in the limit of infinite interaction range, the critical temperature is independent of $\mu$ [see the phase diagram in Fig.~\ref{fig:MF:phase:diagram}(d)]. The finite range of interactions therefore reduces the extension of the oscillating phase in the phase diagram $(T, \mu)$, and it is more difficult, very far from equilibrium, to obtain collective oscillations than in the case of an infinite interaction range.

\subsubsection{Influence of the coupling constant $J_2$}
\begin{figure}
    \centering
    \includegraphics{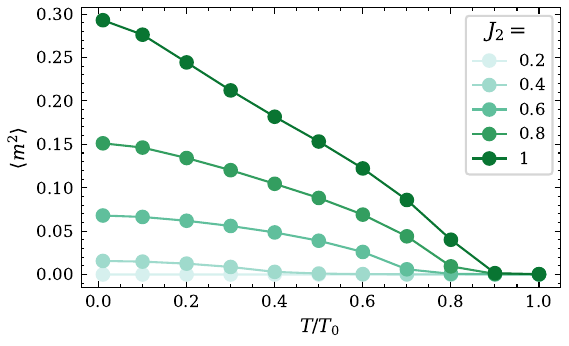}
    \caption{Influence of the ferromagnetic coupling $J_2$ between fields: $\langle m^2\rangle$ as a function of temperature for different values of $J_2$. Parameters: $J_1=0$, $\mu=2$, $a=3$ and $L=31$.  }
    \label{fig:simu:J2}
\end{figure}
We now briefly study the influence of the coupling constant $J_2$ between field variables $h_i$. We take $J_1=0$ [there are no ferromagnetic interactions between the spins]. We plot $\langle m^2\rangle$ as a function of the reduced temperature $T/T_0$ (with $T_0=J_2/2$ when $J_1=0$) for different values of $J_2$ in Fig.~\ref{fig:simu:J2}. We plot the critical temperature $T_c$ for different values of $J_2$ in Fig.~\ref{fig:Tcnumres}(c). We observe that the stronger the ferromagnetic interactions between the fields (higher $J_2$), the more the oscillating phase is extended (the higher the critical temperature is).

\subsubsection{Summary of numerical results} We carried out numerical simulations of the three-dimensional spin model considering short-range interactions, varying the various parameters: $a$, $\mu$, $J_2$. These simulations enabled us to conclude that:
\begin{itemize}
    \item  As the interaction range increases, the oscillating phase becomes more extended (the critical temperature $T_c$ approaches $T_0$), and the oscillating phase appears to have a critical behavior similar to that of noisy interacting oscillators. The oscillating phase seems to disappear when the interaction range is very small. 
    \item  The greater the non-reciprocity parameter $\mu$ (also controlling the distance to equilibrium), the less extended the oscillating phase. The oscillating phase seems to disappear very far from equilibrium.
    \item The lower the ferromagnetic interactions between the fields (controlled by $J_2$), the less extended the oscillating phase, and the smaller the oscillation amplitude. The oscillating phase seems to disappear when $J_2$ is low. 
\end{itemize}
We now seek to understand these behaviors using analytical tools, by linking them to a model of interacting noisy oscillators.

\section{Link with the synchronization transition of noisy oscillators}
\label{sec:link:synchro}

\subsection{Effective model of interacting noisy oscillators}
\label{sec:effective:model}

In order to make the link between the microscopic model of spins and fields, and a model where oscillators would appear at the mesoscopic level, we introduce a local averaging procedure. To do this, we divide space into distinct boxes of size $n_b=(2\lfloor{a}\rfloor+1)^3$. 
We denote as $N_G=N/n_b$ the total number of boxes. 
We relabel the spins and the fields as follows: the boxes are indexed with an index $k$ and within each box, the $n_b$ sites are indexed by a position $i$ with $i=0$ the central position. 
We then denote $\sigma_k^1(i)$ the spins and $\sigma_k^2(i)$ the fields.
The division into boxes can be done in different ways, depending on the choice of central spins, and we consider an arbitrary choice of central spins. 
We denote $M_k(i)$ the average magnetization of spins separated by a distance less than $a$ from the spin of the box $k$ at position $i$:
\be
M_k(i)=\frac{1}{n(a)}\sum_{d(({k'},{i'}),(k,i))\leq a}  
\sigma_{k'}^1(i'),
\ee
where $\sum_{d(({k'},{i'}), (k,i))\leq a}$ corresponds to a sum over $k'$ and $i'$ such that the site of box $k'$ at position $i'$ is located at a distance less than $a$ from the site of box $k$ at position $i$. 
We denote as
\be M_k=M_k(0)\ee
the central spin of box $k$. We also introduce $H_k(i)$ as
\be
H_k(i)=\frac{1}{n(a)}\sum_{d(({k'},{i'}), (k,i))\leq a} \sigma_{k'}^2(i'),
\ee
and $H_k=H_k(0)$ the local averages of the fields. 

We now seek to describe the system only on the $2^{2N_G}$ configurations $(\vec{M}, \vec{H})=(M_1, ..., M_{N_G}, H_1, . ..., H_{N_G})$ instead of the $2^{2N}$ configurations of spins and fields $(\vec{s}, \vec{h})=(s_1, ..., s_N, h_1, ..., h_N)$.
We therefore introduce the probability $P_{N_G}(\vec{M}, \vec{H})$ defined as
\be \label{eq:def:pNM1H1...} P_{N_G}(\vec{M}, \vec{H})=\sum_{(\vec{s}, \vec{h})\in S(\vec{M}, \vec{H})}P(\vec{s},\vec{h})\ee
where $S(\vec{M}, \vec{H})$ represents the set of configurations with local magnetization $M_k$ and local field $H_k$ for all groups $k\in [1, N_G]$.
Calculations are detailed in \ref{app:FP}, and we now explain the main steps of the reasoning. 
To obtain an equation on $P_{N_G}(\vec{M}, \vec{H})$, we inject Eq.~(\ref{eq:def:pNM1H1...}) into the microscopic master equation 
governing the evolution of the probability of configurations $(\vec{s}, \vec{h})$.
We make the assumption that $n(a)\gg 1$ (the equations obtained will therefore not be valid for small $a$) in order to perform a perturbative expansion in $1/n(a)$ and keep only terms up to second order in this expansion. 
Next, we assume that in the same box $k$, the values of the local magnetization (and field) are all comparable, namely $M_k(i)\approx M_k$ and $H_k(i)\approx H_k$.
For the zeroth-order term in $M_k(i)-M_k$ and $H_k(i)-H_k$, all spins (and fields) in box $k$ have identical contributions, as the transition rates depend only on $M_k(i)$ and $H_k(i)$. The zeroth-order term therefore gives for each individual box $k$ a contribution similar to that of the mean-field spin model. The sums of the first-order terms in $M_k(i)-M_k$ and in $H_k(i)-H_k$ can be approximated by Laplacian terms, proportional to an interaction constant $C$. Finally, as we wish to study the onset of oscillations, we also assume that $M_k$ and $H_k$ are small, in order to simplify the resulting expressions. 
We then obtain the following equation on $P_{N_G}(\vec{M}, \vec{H})$:
\be \label{eq:FP:res}\begin{split}\partial_t  &P_{N_G}(\vec{M}, \vec{H})=\\
\sum_k \Bigg[&\partial_{M_k}\left(M_k-\tanh\left[\frac{J_1M_k+H_k}{T}\right]-\frac{C}{T}\Delta (J_1M_k+H_k)\right) P_{N_G}(\vec{M}, \vec{H})\\
+&\partial_{H_k}\left(H_k-\tanh\left[\frac{J_2H_k+(1-\mu)M_k}{T}\right]-\frac{C}{T}\Delta (J_2H_k+(1-\mu)M_k)\right) P_{N_G}(\vec{M}, \vec{H})\\
+&\frac{1}{n(a)}\partial_{M_k}^2 P_{N_G}(\vec{M}, \vec{H})+\frac{1}{n(a)}\partial_{H_k}^2P(\vec{M}, \vec{H})\Bigg].\end{split}\ee
where $\Delta$ is the Laplacian.
Eq.~(\ref{eq:FP:res}) corresponds to a Fokker-Planck equation on $P_{N_G}(\vec{M}, \vec{H})$, from which we can write the associated Langevin equations:
    \be\label{eq:langevin} \begin{split}\frac{dM_k}{dt}& =-M_k+\tanh[(J_1M_k+H_k)/T]+\frac{C}{T}\Delta(J_1M_k+ H_k)+\sqrt{\frac{2}{n(a)}}\xi_{k}^1, \\
\frac{dH_k}{dt} &=-H_k+\tanh[(J_2H_k+(1-\mu)M_k)/T]+\frac{C}{T}\Delta [J_2H_k+(1-\mu)M_k]+\sqrt{\frac{2}{n(a)}}\xi^{2}_k,\end{split}\ee
with $\xi^i_k$ independent Gaussian white noises with correlations
\be \langle \xi^l_k( t) \xi^{l'}_{k'}( t')\rangle =\delta_{k, k'}\delta_{l, l'}\delta(t-t').\ee
These equations are similar to those describing interacting noisy oscillators \cite{risler.etal.2004}. Without interaction ($C=0$), each oscillator corresponds to an oscillator obtained from a mean-field model of $n(a)$ spins and fields, which therefore exhibits oscillations for $T<T_0$ and $\mu>1+(J_1-J_2)^2/4$.  Each oscillator is subject to a noise whose amplitude is proportional to $1/\sqrt{n(a)}$. For $C\neq 0$, oscillators interact with their nearest neighbors. 

To understand the influence of each parameter, we derive the normal form of these equations, obtained for $M_k, H_k\ll 1$. We introduce
\be Z_k=(-1+J_1/T)M_k+H_k/T+i\sqrt{v_0}M_k,\ee 
with 
\be \label{def::v0} v_0=\frac{1}{T}\sqrt{\mu-1-\frac{(J_1-J_2)^2}{4}}.\ee
We change the angular reference frame to suppress the average frequency dependence, by defining $\tilde{Z_k}(t)=e^{-i\sqrt{v_0}t}Z_k(t)$.
We neglect the imaginary part of the coefficients of the terms in $|\tilde{Z_k}|^2\tilde{Z_k}$ and in $\Delta \tilde{Z_k}$ as they contribute to a change of frequency, which is small compared to the mean frequency $\sqrt{v_0}$. We find the following Langevin equations on $\tilde{Z_k}$:
\be \label{eq:Ztilde}\frac{d\tilde{Z}_k}{dt}=(T_0/T-1)\tilde{Z}_k-U|\tilde{Z}_k|^2\tilde{Z}_k+\frac{T_0C}{T}\Delta \tilde{Z}_k+\sqrt{\frac{\alpha}{n(a)}}\left[\eta_k^R+i\eta_k^I\right]\ee
with $T_0$ defined in Eq.~(\ref{def:T0}), and
\bea
\label{def:alpha} &&\alpha=\frac{1}{T^2}\left[(J_1-T)^2+1+(\mu-1)-\frac{(J_1-J_2)^2}{4}\right],\\
\label{def:U}&&U(T=T_0)=\frac{(J_1+(\mu-1)J_2)(\mu-1+J_1J_2)}{(J_1+J_2)[4(\mu-1)-(J_1-J_2)^2]}.
\eea
In Eq.~(\ref{eq:Ztilde}), $\eta_k^R$ and $\eta_k^I$ are independent Gaussian white noises, satisfying $\langle \eta_k^R(t) \eta_{k'}^R(t')\rangle=\langle \eta_k^I(t)\eta_{k'}^I(t')\rangle =\delta_{k, k'}\delta(t-t')$ and $\langle \eta_k^R\eta_k^I\rangle=0$.
To reduce the number of parameters involved in these equations, we reabsorb the parameter $U$ into a new variable $z_k$ defined as
\be z_k=\sqrt{U}\tilde{Z}_k.\ee
We further introduce the two parameters:
\be \label{def:D} D=\frac{U\alpha}{n(a)}\ee
and 
\be \label{def:gamma}\gamma=T_0/T,\ee
so that Eq.~(\ref{eq:Ztilde}) becomes
\be \label{eq:zk:final}\frac{dz_k}{dt}=(\gamma-1)z_k-|z_k|^2z_k+\gamma C\Delta z_k+\sqrt{D}\left[\eta_k^R+i\eta_k^I\right].\ee
The parameter $C$ can be estimated from the microscopic model [see \ref{app:FP}], and is found to depend only weakly on the value of interaction range $a$,
at odds with the parameters $D$ and $\gamma$.

We now discuss the properties of Eq.~(\ref{eq:zk:final}) for generic values of the coefficients, temporarily ignoring the relation to the microscopic model.
The link between this equation and the microscopic model will then be discussed in section~\ref{sec:comparison}. 
Eq.~(\ref{eq:zk:final}) corresponds to a stochastic Ginzburg-Landau equation \cite{risler.etal.2005,aranson.rmp2002}, which describes interacting noisy oscillators. 
The parameter $D$ controls the noise amplitude. 
The parameter $\gamma$ controls the amplitude of the oscillations. Indeed, when the oscillators are neither noisy ($D=0$) nor interacting ($C=0$), then oscillations of amplitude $z_k=\sqrt{\gamma-1}$ appear for $\gamma\geq 1$ on each oscillator. 
Finally, the parameter $C$ controls the amplitude of the interactions. For $C=0$, the oscillators do not interact and are thus unsynchronized. 
In models of noisy interacting oscillators, the noise is often neglected on the amplitude of the oscillations, allowing the equations to be reduced to a phase equation for constant-amplitude oscillators \cite{acebron_kuramoto_2005}.  However, in the model presented here, the oscillation amplitude may be small compared to the noise amplitude ($\gamma-1\ll D$),
so that the assumption of constant-amplitude oscillators is no longer valid.

\subsection{Phase diagram of the synchronization of noisy oscillators in the mean field limit}
\label{sec:MFnoisy:oscill}

\subsubsection{Mean field limit}
Our goal is now to study the mean-field phase diagram of the coupled noisy oscillators described by Eq.~(\ref{eq:zk:final}).
While the parameters $D$ and $\gamma$ have significant variations as a function of microscopic parameters of the three-dimensional spin model (like the interaction range $a$),
the parameter $C$ remains almost constant with a value $C\approx 0.025$ [see \ref{app:FP}].
We thus take $C$ as a constant in the following, and focus on the influence of the parameters $D$ and $\gamma$ on the behavior of Eq.~(\ref{eq:zk:final}).
As a mean-field approximation to the three-dimensional model described in Eq.~(\ref{eq:zk:final}), we consider the solvable case where all the individual noisy oscillators interact with each other in a mean-field way: 
\be \label{eq:MF:approx}
\Delta z_k 
\to 6(\overline{z}-z_k)
\ee
where $6$ corresponds to the number of nearest neighbors in a cubic lattice in dimension $d=3$ and
\be \overline{z}=\frac{1}{N_G}\sum_{k=1}^{N_G} z_k\ee 
is the mean value of $z_k$ over all boxes.
To lighten notations, we introduce $c=6C\approx 0.15$.
In polar coordinates, we write:
\be \overline{z}=Re^{i\Theta},\quad  z=r e^{i\theta}.\ee
We take the limit $N_G\to\infty$, and since we are considering the limit where all boxes interact, we introduce $P_1(x, y)$, the distribution for a single box, such that we assume
\be P_{N_G}(\{r_1, \theta_1, ..., r_{N_G}, \theta_{N_G}\})=\prod_{i=1}^N P_1(r_i, \theta_i).\ee
We can then obtain $P_1(r, \theta)$ as [see \ref{app:mean:field:complement} for detailed calculations]:
\be\label{eq:p1rtheta} P_1(r, \theta)=p_0 \exp {\Bigg[-\frac{2}{D}\left(\frac{1}{4}r^4-\frac{(\gamma-1)}{2}r^2+\frac{\gamma c}{4}\left(R^2+r^2-2Rr\cos(\Theta-\theta)\right)\right)\Bigg]},\ee
where $p_0$ is a normalization constant such that
\be \int dr r d\theta P_1(r, \theta)=1.\ee
The values of $R$ and $\Theta$ are determined from the following self-consistent equation:
\be\label{eq:self:consistent:2} Re^{i\Theta}=\frac{\int_{0}^{\infty} dr \int_{-\pi}^{\pi} d\theta r^2e^{i\theta} \exp\left[{-\frac{2}{D}\left(\frac{1}{4}r^4-\frac{\gamma-1}{2}r^2+\frac{\gamma c}{4}\left(r^2+R^2-2Rr\cos(\Theta-\theta)\right)\right)}\right]}{\int_{0}^{\infty} dr \int_{-\pi}^{\pi} d\theta r \exp\left[{-\frac{2}{D}\left(\frac{1}{4}r^4-\frac{\gamma-1}{2}r^2+\frac{\gamma c}{4}\left(r^2+R^2-2Rr\cos(\Theta-\theta)\right)\right)}\right]}\ee 
which comes from the definition: $\overline{z}=Re^{i\Theta}=\int r dr d\theta re^{i\theta}P_1(r,\theta)$.
A change of variable $\theta'=\theta-\Theta$ in the integrals shows that Eq.~(\ref{eq:self:consistent:2}) is independent of $\Theta$. The global phase $\Theta$ is therefore arbitrary, and only the relative phase $\theta-\Theta$ indicates the presence or absence of synchronization.

\begin{figure}
    \centering
    \includegraphics{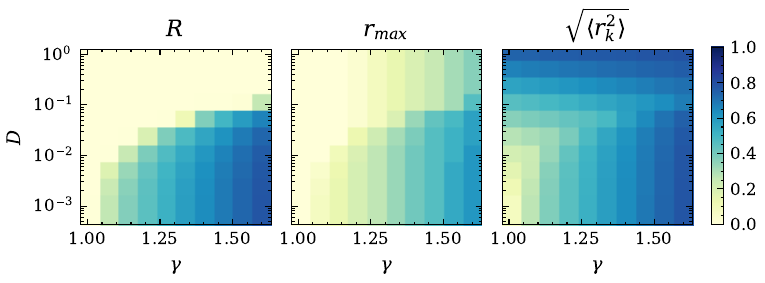}
    \caption{Phase diagram of the noisy coupled oscillator model in the mean-field limit. Amplitudes $R$ (left panel), $r_{max}$ (middle panel) and $\sqrt{\langle r_k^2\rangle}$ (right panel) in the $(\gamma, D)$ plane obtained numerically from Eqs.~(\ref{eq:p1rtheta}) and (\ref{eq:self:consistent:2}).  }
    \label{fig:MF:R:r2}
\end{figure}
\begin{figure}
    \centering
    \includegraphics{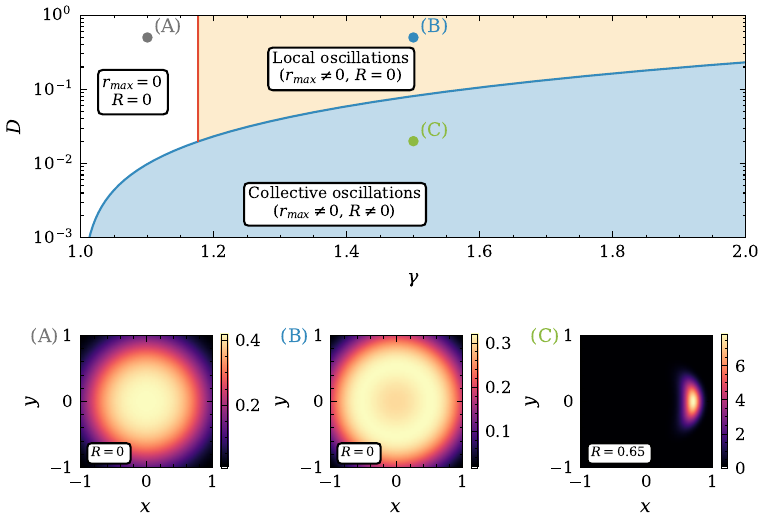}
    \caption{Top: Phase diagram of the noisy coupled oscillator model in the mean-field limit. The blue curve corresponds to $\gamma_c(D)$ [see Eq.~(\ref{eq:MF:solution})], the red curve corresponds to $\gamma_{loc}$ [see Eq.~(\ref{eq:gamma:loc})]. 
    Three different phases are observed: a phase with global oscillations where $r_{max}\neq 0$ and $R\neq0$, a phase with unsynchronized local oscillations where $r_{max}\neq0$ and $R=0$, and a phase with noise $r_{max}=0$ and $R=0$. Bottom: Examples of probability distributions $P_1(r, \theta)$, plotted with coordinates $x=r\cos(\theta-\Theta)$ and $y=r\sin(\theta-\Theta)$, for each region of the phase diagram, marked by (A), (B) and (C) in the top panel.
    }
    \label{fig:oscillateurs:MF:phase:diagram}
\end{figure}

\subsubsection{Presence of collective oscillations}
Since $R$ corresponds to the amplitude of $\overline{z}$, there is a transition to (partially) synchronized oscillations (with frequency $\sqrt{v_0}$) when
Eq.~(\ref{eq:self:consistent:2}) admits a solution with $R\neq 0$.
In Fig.~\ref{fig:MF:R:r2}, we show with a colorcode in the $(\gamma, D)$-plane the value of $R$ ($0 \le R \le 1$), obtained from Eq.~(\ref{eq:self:consistent:2}).
We obtain that for high values of noise $D$ and low values of $\gamma$, there are no macroscopic oscillations, $R=0$, while such oscillations are present for low values of noise. 
Expanding Eq.~(\ref{eq:self:consistent:2}) for $R\ll 1$, we find that a non-zero $R$ solution appears when the pair $(\gamma, D)$ verifies the following equation:
\be\begin{split}\label{eq:MF:solution} 1=\frac{\gamma c}{D}\Bigg[\gamma(1-c)-1
    +\frac{\sqrt{\frac{2 D }{\pi }}\exp\left[{-\frac{(\gamma(1- c) -1)^2}{2D}}\right]}{\text{erfc}\left(\frac{ -\gamma (1-c)+1}{\sqrt{2D }}\right)}\Bigg]\end{split}\ee
where $\text{erfc}(x)=1-\text{erf}(x)$, and $\text{erf}(x)$ is the error function. 
This equation can be solved numerically to obtain the critical parameters $\gamma_c(D)$ (or $D_c(\gamma)$) where a solution $R\neq 0$ appears. The results of which are shown in Fig.~\ref{fig:oscillateurs:MF:phase:diagram}.
Taking the limit $D\ll 1$ and $\gamma\sim 1$, we obtain 
\be\label{res:gammac} \gamma_c=1+\frac{2D}{c}.\ee
In the low-noise limit ($D\to0$), we obtain $\gamma_c=1$ when $c\neq0$: as soon as oscillations appear locally ($\gamma=1$), they are synchronized.

Models of oscillator synchronization transitions often start with oscillators that oscillate at a given frequency and amplitude, usually with a small noise compared to the oscillation amplitude. These models can then be used to study whether the oscillators are synchronized or not  \cite{acebron_kuramoto_2005} by neglecting the noise on the oscillator amplitude. In the model introduced here, the oscillation amplitude can become negligible compared to the noise, which adds a third phase to the phase diagram, with neither collective nor local oscillations. We now turn to the condition for oscillations to occur locally, but to be desynchronized on a global scale.

\subsubsection{Presence of local oscillators}
We try here to determine the parameter values for which oscillators exist locally, but are unsynchronized on a global scale. First, we need to choose the criterion for the presence of local oscillations. Oscillations are considered to exist when the maximum of $P_1(r, \theta)$ occurs at a non-zero value of $r$, denoted $r_{max}$. In Fig.~\ref{fig:MF:R:r2}, we show in color the value of $r_{max}$ in the plane $(\gamma, D)$ obtained numerically from equations (\ref{eq:p1rtheta}) and (\ref{eq:self:consistent:2}). And from Eq.~(\ref{eq:p1rtheta}), we obtain that local oscillations appear for
$\gamma=\gamma_{loc}$, with
\be \label{eq:gamma:loc}
\gamma_{loc} = (1-c)^{-1}.
\ee
For low values of $\gamma$ (i.e., $\gamma<\gamma_{loc}$), there are no local oscillations. 
The interaction amplitude $c$ is therefore the quantity that controls the extent to which oscillators can exhibit local oscillations. As $c$ is fixed in the model, we cannot control the onset of local oscillations. 
The $\gamma_{loc}$ curve, which is independent of $D$, is plotted in Fig.~\ref{fig:oscillateurs:MF:phase:diagram} in the ($\gamma, D$) plane.

\subsubsection{Summary of the phase diagram}
To sum up, we have obtained a phase diagram in the space $(\gamma, D)$ shown in Fig.~\ref{fig:oscillateurs:MF:phase:diagram}. We observe a phase where there are global oscillations ($R\neq0$) for small values of $D$ and give an example of the probability density $P_1(x, y)$ in Fig.~\ref{fig:oscillateurs:MF:phase:diagram}(C). The average phase of each oscillator is $\theta_k=\Theta$, and there is little noise in both amplitude and phase. 
There is a second phase, for low values of $\gamma$ and high values of $D$, where there are no oscillators locally, and the $z_k$ variables are only subject to noise. An example of $P_1(x, y)$ in this phase is given in Fig.~\ref{fig:oscillateurs:MF:phase:diagram}(A). Finally, there is a phase, for high values of noise $D$ and high values of $\gamma$, where effective local oscillators emerge (i.e., $P_1(x, y)$ is maximal for a non-zero value of $r=\sqrt{x^2+y^2}$) but are not synchronized. In this case, the phase $\theta$ is uniformly distributed and therefore $R=0$. An example of $P_1(x, y)$ in this phase is shown in Fig.~\ref{fig:oscillateurs:MF:phase:diagram}(B).

\subsubsection{Influence of noise on local oscillation detection}
It is sometimes interesting to look only at moments of the probability density $P_1(r,\theta)$, such as the second moment $\langle r_k^2\rangle$. 
In this case, the condition for the existence of local oscillations is different, as noise can hide the presence of very low-amplitude oscillations. Indeed, when $R=0$, we obtain from Eq.~(\ref{eq:p1rtheta}),
\be \langle r_k^2\rangle=\gamma(1-c)-1+\sqrt{\frac{2D}{\pi}}\frac{\exp\left[{-\frac{(\gamma(1-c)-1)^2}{2D}}\right]}{1+\text{erf}\left(\frac{\gamma(1-c)-1}{\sqrt{2D}}\right)}.\ee
We therefore find that if 
\be \gamma(1-c)-1\ll D^{1/2},\ee 
fluctuations dominate and even if $r_{max}\neq 0$, it is difficult to distinguish this case from the one where $P_1(r, \theta)$ is maximal in $r=0$. We calculate $\langle r_k^2\rangle$ from Eq.~(\ref{eq:p1rtheta}) and we represent $\sqrt{\langle r_k^2\rangle}$ in the $(\gamma, D)$-plane in Fig.~\ref{fig:MF:R:r2}. 
The second moment $\langle r_k^2\rangle$ is not sufficient to determine whether local oscillations are present or not. 

\subsection{Comparison with 3D lattice simulations of interacting noisy oscillators }

\begin{figure}
    \centering
    \includegraphics{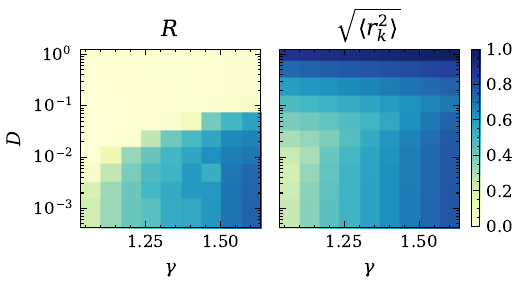}
    \caption{Numerical study of the phase diagram of the three-dimensional noisy coupled oscillator model defined by Eq.~(\ref{eq:zk:final}), simulated on a cubic lattice of size $N=10^3$. Amplitudes $R$ and $\sqrt{\langle r_k^2\rangle}$ are plotted in the ($\gamma, D)$ plane, to detect the presence of global (left panel) or local (right panel) oscillations.}
    \label{fig:3D:simu:noisyoscillators}
\end{figure}

In Sec.~\ref{sec:MFnoisy:oscill}, we considered a mean-field limit where all boxes interact with each other.
We now compare these mean-field results with the behavior of the three-dimensional model of noisy oscillators given in Eq.~(\ref{eq:zk:final}),
which was derived from the three-dimensional spin model. We therefore run simulations of the Langevin equation Eq.~(\ref{eq:zk:final}) on a 3D lattice to compare with the mean-field results. We take $N=1000$ oscillators in the numerical simulations. The amplitude $R$ of $\overline{z}$ and the second moment $\sqrt{\langle r_k^2\rangle}$ are plotted in the $(\gamma, D)$ plane in Fig.~\ref{fig:3D:simu:noisyoscillators}, and we observe a qualitative agreement with results obtained in the mean-field  limit [Fig.~\ref{fig:MF:R:r2}]. The phase without global oscillations corresponds to high $D$ and low $\gamma$, the synchronized phase corresponds to low $D$ and high $\gamma$, and there is a phase with local but unsynchronized oscillators for high $\gamma$ and high $D$ (which can be confirmed by looking at the probability density $P_1(x, y)$ not shown here, insofar as $R$ and $\sqrt{\langle r_k^2\rangle}$ are not enough to identify this phase).

\section{Comparison between the spin model and the interacting noisy oscillator model}
\label{sec:comparison}
We have obtained the phase diagram for interacting noisy oscillators, where the two main control parameters are the noise amplitude $D$ and the parameter
$\gamma$ controlling the amplitude of oscillations in the absence of interactions [see Eqs.~(\ref{def:D}) and (\ref{def:gamma}) for the definition of $D$ and $\gamma$].
We now come back to the three-dimensional spin model, and seek to explain qualitatively the results observed in the simulations by comparing numerical results with the analytical results of the mean-field model of noisy oscillators, taking into account the expression of the parameters $D$ and $\gamma$ in terms of the microscopic parameters of the three-dimensional spin model.

\subsection{Influence of interaction range $a$}
The interaction range $a$ enters only the definition of the parameter $D$, with 
$D\propto n(a)^{-1}$
where $n(a)$ is the number of sites with which a spin or a field interacts. We obtained that $\gamma_c-1\sim 2D/c$ in the mean-field limit of interacting noisy oscillators [see Eq.~(\ref{res:gammac})]. Since $\gamma=T_0/T$, the critical temperature depends on $a$ as 
\be \frac{T_0-T_c}{T_0}\sim \frac{1}{n(a)}.\ee
This is the dependence obtained for stochastic simulations of the three-dimensional spin model [see Fig.~\ref{fig:Tcnumres}(a)]. Qualitatively, when the neighborhood size is large, low-noise oscillators appear locally and synchronize, whereas for small neighborhoods, the noise is too high and oscillations are destroyed.

\subsection{Influence of the non-reciprocity parameter $\mu$}
\begin{figure}
    \centering
    \includegraphics{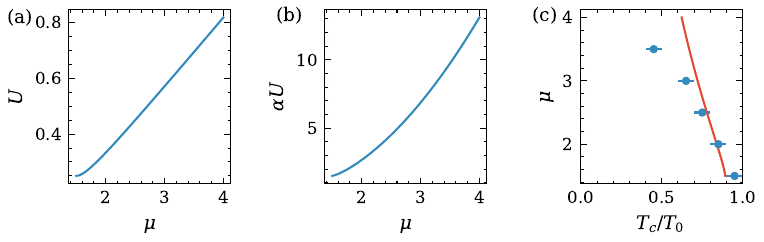}
    \caption{Influence of $\mu$ on parameters (a) $U$ controlling oscillation amplitude and (b) $\alpha U$ controlling noise amplitude ($D\propto \alpha U$). (c) Comparison of critical temperatures $T_c$ normalized by $T_0$ as a function of $\mu$ obtained simulations of the three-dimensional spin model [blue dots, from simulations reported in Fig.~(\ref{fig:simu:mu})], with the noisy coupled oscillator model in the mean-field approximation [red curve, solution of Eq.~(\ref{res:gammac})]. Parameters: $J_1=0$, $J_2=1$ [for (a), (b), (c)] and $a=3$ [for (c)]. }
    \label{fig:D:mu}
\end{figure}
Before comparing critical temperatures, we note a strong dependence of $m$ and $\dot{m}$ with $\mu$. The equation that controls the behavior of interacting noisy oscillators is Eq.~(\ref{eq:zk:final}). To arrive at this equation, we started with the variables $M_k$ and $H_k$, corresponding to the local magnetization and the local mean field, and made the following change of variable:
\be \label{eq:chgt:var:inverse}
(-1+J_1/T)M_k+H_k/T+i\sqrt{v_0}M_k=e^{i\sqrt{v_0}t}z_k/\sqrt{U},\ee
with $v_0=\sqrt{\mu-\mu_0}/T$, where $\mu_0(T_0)=1+(J_1-J_2)^2/4$.
In terms of the variable $z_k$ (or $\overline{z}$) we have 
\be M_k \sim z_k/\sqrt{v_0U},\quad m\sim \overline{z}/\sqrt{v_0U}.\ee 
The variable $\dot{m}$ defined in Eq.~(\ref{def:mdot}), and corresponding to the stochastic derivative of the magnetization, can be rewritten as
\be\dot{m}= \frac{1}{N_G}\sum_{k=1}^{N_G}\sum_{i=1}^{n_{b}} \frac{-2\sigma_k^l(i)}{n_b}\frac{1}{1+\exp\left[-2\sigma_k^l(i)[J_1M_k(i)+H_k(i)]/T\right]}.\ee
Keeping only the lowest order in $M_k(i)-M_k$ and $H_k(i)-H_k$, we then obtain the following approximation for $\dot{m}$:
\be
\dot{m}=\frac{1}{N_G}\sum_{k=1}^{N_G} \left[-M_k+\tanh\left(\frac{J_1M_k+H_k}{T}\right)\right].
\ee
At lowest order in $M_k$ and $H_k$, we get $\dot{m}\approx (-1+J_1/T)m+h/T$, which leads to the following scaling law [see Eq.~(\ref{eq:chgt:var:inverse})]
\be \dot{m}\sim \overline{z}/\sqrt{U}.\ee
The dependence of $U$ on $\mu$ is given in Eq.~(\ref{def:U})(a). We plot $U$ as a function of $\mu$ in Fig.~\ref{fig:D:mu}(a) for $J_1=0$ and $J_2=1$, and obtain that $U$ increases with $\mu$. 
As $m\sim z/\sqrt{v_0U}$ we therefore obtain that $m\sim z/\mu^2$ (neglecting $\mu_0$ with respect to $\mu$), and $\dot{m}\sim z/\mu$. We can therefore explain the trend observed in Fig.~\ref{fig:simu:mu}: $\langle m^2\rangle$ and $\langle\dot{m}^2\rangle$ decrease with $\mu$, but the decrease in $\langle m^2\rangle$ is much stronger. This explains why, for high values of $\mu$, $m$ and $\dot{m}$ are very small, making it difficult to observe oscillations numerically for finite system sizes.

For values of $\mu$ up to $\mu\approx 4$, we were still able to measure critical temperatures [see Fig.~\ref{fig:simu:mu}]. We obtain that $T_c$ decreases when $U$ increases. To understand this behavior, we determine the dependence of the coefficient $D$ on $\mu$. We have 
\be D\propto \alpha U,\ee
where $\alpha$ and $U$ are given in Eqs.~(\ref{def:alpha}) and (\ref{def:U}). The dependence of $\alpha U$ is shown in Fig.~\ref{fig:D:mu}(b) for $J_1=0$ and $J_2=1$. Except for low values of $\mu$ (where the oscillating phase no longer exists), $\alpha U$ and therefore the noise parameter $D$ increase with $\mu$. According to the phase diagram shown in Fig.~\ref{fig:oscillateurs:MF:phase:diagram}, as the noise parameter $D$ increases, the transition takes place at higher $\gamma=T_0/T$ and therefore the critical temperature $T_c$ decreases. In Fig.~\ref{fig:D:mu}(c), we plot the evolution of $T_c$ for different values of $\mu$. The points correspond to stochastic simulations of the three-dimensional spin model [results shown in Fig.~\ref{fig:simu:mu}] and the curve corresponds to $T_c(\mu)$ obtained for the noisy oscillator model with the mean-field couplings [see Eq.~(\ref{res:gammac})]. A quantitative difference between the two models is observed, but the overall trend is similar.

\subsection{Influence de $J_2$}
\begin{figure}
    \centering
    \includegraphics{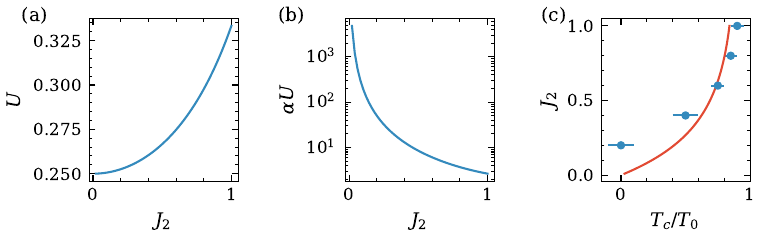}
    \caption{Influence of $J_2$ on parameters (a) $U$ controlling oscillation amplitude and (b) $\alpha U$ controlling noise amplitude ($D\propto \alpha U)$. (c) Comparison of critical temperatures $T_c$ normalized by $T_0$ as a function of $J_2$ obtained from simulations of the three-dimensional spin model [blue dots, from simulations reported in Fig.~(\ref{fig:simu:J2})] with the noisy coupled oscillator model in the mean-field approximation [red curve, solution of Eq.~(\ref{res:gammac})]. Parameters: $\mu=2$, $J_1=0$ [panels (a), (b) and (c)] and $a=3$ [panel (c)]. }
    \label{fig:D:J2}
\end{figure}
We finally study the influence of the amplitude of the ferromagnetic interactions between the fields $J_2$ (for $J_1=0$). For the spin model simulations, we observed two main features: $\langle m^2\rangle$ increases with $J_2$ and the transition temperature also increases with $J_2$.  
For the model of noisy oscillators interacting with mean-field couplings, the coefficient $U$ depends only weakly on $J_2$ [see Fig.~\ref{fig:D:J2}(a)],
so that the value of $\dot{m}$ changes only slightly if the oscillators are completely synchronized. However, $m\sim (J_1+J_2)\overline{z}/\sqrt{U}$ as $v_0\propto 1/(J_1+J_2)^2$ and therefore $m$ depends strongly on the value of $J_2$. Furthermore, we find that the coefficient $D\propto \alpha U$ increases as $J_2$ increases [see Fig.~\ref{fig:D:J2}(b)], which implies that the transition temperature increases as $J_2$ increases. In Fig.~\ref{fig:D:J2}(c), we plot the scaled critical temperature $T_c/T_0$ obtained from Eq.~(\ref{eq:MF:solution}) for different values of $J_2$
(with $J_1=0$, $a=3$ and $\mu=2$) and compare with the critical temperatures obtained from simulations of the three-dimensional spin model
with the same parameters. A qualitative agreement between the two models is observed.

The influence of the ferromagnetic interaction $J_1$ between the spins is briefly discussed in \ref{app:J1}. We obtain that the coupling constant $J_1$ has a similar role as the coupling constant $J_2$.

\section{Conclusion}

We have considered a stochastic model consisting of spins and fields arranged on a three-dimensional cubic lattice, with interactions on a spherical neighborhood. We studied the influence of the interaction range
using numerical simulations and analytical tools. We have shown that decreasing the range of ferromagnetic couplings, or strongly increasing the non-reciprocity of on-site interactions between spins and fields, hinders the onset of collective oscillations.
To understand the phase diagram of the model in terms of microscopic parameters, we linked the three-dimensional spin model, in which individual degrees of freedom do not oscillate in the absence of interactions, to an effective model of coupled noisy oscillators on a cubic lattice, which we analyzed numerically as well as analytically within a mean-field approximation.
This mapping allowed us to propose the following qualitative interpretation of the transition scenario in three dimensions. At the local level,
noisy oscillations emerge (similar to the mean-field case, but for a finite-size system), whose amplitude and fluctuations are related
to the interaction range. Subsequently, these local oscillators may synchronize at system level.
The coherence of the emerging picture a posteriori justifies the approximations made in the analytical derivations.
 
As for future work, it would be useful to study such finite-dimensional systems more in depth, notably as a function of space dimension. 
One may expect that the onset of local noisy oscillations combined with a large-scale synchronization transition may be a generic two-stage scenario for the onset of collective oscillations in finite-dimensional spin system, that may be worth confirming in different models and in different dimensions.
However, this scenario is likely to be modified in low dimensions. In particular, for two-dimensional systems,
the oscillating phase has been shown to be destroyed by the propagation of defects \cite{avni.etal.2023}.
In addition, it would also be of interest to develop a renormalization group approach starting from the spin level of description,
thereby extending the results of \cite{risler.etal.2004,risler.etal.2005} based on the synchronization of oscillators.



\appendix

\section{Simulation methods}

\subsection{Simulations with nearest-neighbor interactions}

\label{app:simu:a1}
\begin{figure}
    \centering
    \includegraphics[width=1\linewidth]{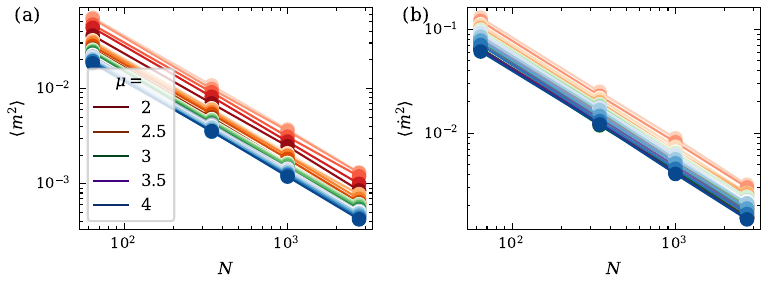}
    \caption{Order parameters (a) $\langle m^2\rangle$ and (b) $\langle\dot{m}^2\rangle$ obtained from numerical simulations of the spin model on a cubic lattice in three dimensions with nearest-neighbor interactions ($a=1$) for different $\mu$ (different colors indicated in the legend) and different temperature $T$ (different transparencies, $T\in[0. 01, 0.05, 0.1, 0.15, 0.2, 0.25, 0.3, 0.35, 0.4, 0.45, 0.5, 0.55]$ from lightest to darkest). Parameters: $J_1=0$ and $J_2=1$.}
    \label{app:fig:a1}
\end{figure}
Here we present results of simulations of the three-dimensional spin model on a cubic lattice, with nearest-neighbor interactions ($a=1$), to illustrate the results presented in Sec.~\ref{subsubsec:a1}, indicating that for $a=1$ we observe no phase transition. In Fig.~\ref{app:fig:a1}, we plot $\langle m^2\rangle$ and $\langle \dot{m}^2\rangle$ as a function of system size $N$ for different values of the parameters $T$ and $\mu$. As a reminder, in the presence of oscillations, $\langle m^2\rangle$ and $\langle \dot{m}^2\rangle$ go to a nonzero limit when $N\to\infty$.
We plot them for different values of $\mu$ (different colors) and for different temperatures $T$ (different transparency). For all parameters explored, $\langle m^2\rangle$ and $\langle \dot{m}^2\rangle$ decrease in $1/N$, indicating that there is no oscillating phase.

\subsection{Comment on the low-temperature phase close to equilibrium}
\label{app:discussion:mu:petit}
In this article, we do not study lower values of $\mu$ for which a ferromagnetic phase would exist in the long-range interaction limit ($a\to\infty$). According to \cite{avni.etal.2023}, the presence or absence of the finite-dimensional ferromagnetic phase in a model of non-equilibrium spins with ferromagnetic interactions is not clearly identified. At finite sizes, the system may remain locked in an ordered phase. Simulations of very large system sizes would be needed to see that this phase is not stable. Our simulations do not provide a clear-cut answer to this question, so that we do not present numerical results for low $\mu$ values of the three-dimensional model. 

\subsection{Identifying the transition temperature as $\mu$ varies}
\label{app:eval:tc:mu:varie}

\begin{figure}[ht]
    \centering
    \includegraphics[width=1\linewidth]{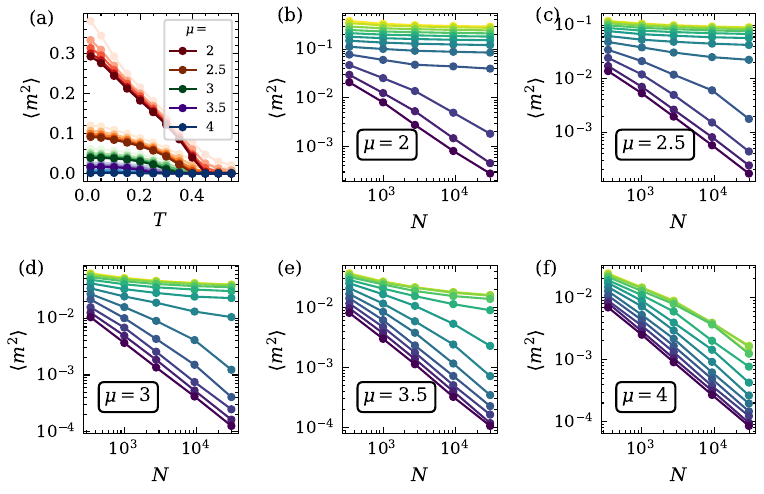}
    \caption{Numerical simulations of the spin model on a three-dimensional lattice with interaction range $a=3$ for different values of $\mu$ and temperatures $T$. (a) $\langle m^2\rangle$ as a function of temperature for different values of $\mu$ (different colors shown in legend) for different system sizes $N=L^3$, $L\in [7, 10, 14, 21, 31]$ from most transparent to least transparent. (b)-(f) $\langle m^2\rangle$ as a function of system size $N$ for different temperatures $T\in[0.01, 0.05, 0.1, 0.15, 0.2, 0.25, 0.3, 0.35, 0.4, 0.45, 0.5, 0.55]$ from yellow/light to violet/dark. Each panel corresponds to a different $\mu$, as indicated. Parameters: $J_1=0$ and $J_2=1$.}
    \label{APP:fig:annexe:mu}
\end{figure}
We explain here how the transition temperature $T_c(\mu)$ was obtained. In Fig.~\ref{fig:simu:mu}, we presented $\langle m^2\rangle$ and $\langle \dot{m}^2\rangle$ as a function of temperature for $L=14$. The magnetization is less numerically expensive than the stochastic derivative of the magnetization, hence to determine the critical temperature $T_c(\mu)$ we use only $\langle m^2\rangle$. In Fig.~\ref{APP:fig:annexe:mu}(a), we plot $\langle m^2\rangle$ as a function of temperature $T$ for different values of $\mu$ (different colors) and different system sizes $N$ (different transparency). To obtain the critical temperature, we plot $\langle m^2\rangle$ as a function of system size $N$ for different temperatures $T$ at fixed $\mu$ [see Figs.~\ref{APP:fig:annexe:mu}(b)-(f)]. When $\langle m^2\rangle$ is constant with $N$, oscillations are presence (since we have seen, for lower system sizes, that $\langle \dot{m}^2\rangle\neq0$) whereas $\langle m^2\rangle\propto1/N$ corresponds to the paramagnetic phase. The critical temperature $T_c$ is the temperature separating these two regimes.

\section{Analytical methods}

\subsection{Details of the derivation of a Fokker-Planck equation}
\label{app:FP}
In this appendix, we detail the reasoning and calculations that led to Eq.~(\ref{eq:FP:res}), a Fokker-Planck equation on the probability density of local, coarse-grained magnetizations $M_k$ and fields $H_k$. The reader is referred to Sec.~\ref{sec:effective:model} for notation definitions.
%
%
We change the notation of the transition rates to take into account the position in the box, $W_s(i)$ becomes
\be \label{FINIE_annexe:taux1}w_1(k, i, \sigma_k^1(i))=\left[1+e^{2\sigma_k^1(i)[J_1M_k(i)+H_k(i)]/T}\right]^{-1}\ee
and $W_h(i)$ turns into
\be \label{FINIE_annexe:taux2}w_2(k, i, \sigma^2_k(i))=\left[1+e^{2\sigma_k^2(i)[J_2H_k(i)+(1-\mu)M_k(i)]/T}\right]^{-1}.\ee

We aim to describe the system only in the $2^{2N_G}$ configurations of $M_k$ and $H_k$ instead of the $2^{2N}$ configurations of the spins and fields $\sigma_k^l$. We write  
\be \vec{\Sigma}=(M_1, ..., M_{N_G}, H_1, ..., H_{N_G})\ee
and $\vec{\sigma}$ is a configuration of all spins $\sigma_k^1(i)$ and fields $\sigma_k^2(i)$. 
We introduce 
\be \label{APP:FINIE:eq:def:pNM1H1...} P_{N_G}(\vec{\Sigma})=\sum_{\vec{\sigma}\in \mathcal{S}(\vec{\Sigma})}P(\vec{\sigma})\ee
where $\mathcal{S}(\vec{\Sigma})$ represents the set of configurations with local magnetization $M_k$ and local field $H_k$ for each group $k\in [1, N_G]$.
The master equation for microscopic $\vec{\sigma}$ configurations reads as follows
\be\label{FINIE_annexe:eq:Psigma} d_tP(\vec{\sigma})=\sum_{\vec{\sigma}'(\neq \vec{\sigma})}[W(\vec{\sigma}|\vec{\sigma}')P(\vec{\sigma}')-W(\vec{\sigma}'|\vec{\sigma})P(\vec{\sigma})].\ee
Using the new notations (\ref{FINIE_annexe:taux1}) and (\ref{FINIE_annexe:taux2}) for the transition rates, Eq.~(\ref{FINIE_annexe:eq:Psigma}) can be rewritten as
\be d_t P(\vec{\sigma})=\sum_{l=1, 2}\sum_{k=1}^{N_G}\sum_{i=0}^{n_b-1}\left[w_l\left(k, i, -\sigma^l_k(i)\right)P(\vec{\sigma}^l_k(i))-w_l(k, i, \sigma^l_k(i))P(\vec{\sigma})\right],\ee
where $\vec{\sigma}^l_k(i)$ is the configuration obtained from $\vec{\sigma}$ after flipping the spin ($l=1$) or field ($l=2$) at position $i$ in box $k$. 
We introduce a sum over $\sigma=\pm1$ to replace the spin $\sigma_k^l(i)$ in the transition rates, also introducing the corresponding Kronecker deltas:
\be d_t P(\vec{\sigma})=\sum_{l=1, 2}\sum_{k=1}^{N_G}\sum_{i=0}^{n_b-1}\sum_{\sigma=\pm1}w_l(k, i, \sigma) \left[\delta(\sigma+\sigma^l_k(i))P(\vec{\sigma}^l_k(i))-\delta(\sigma-\sigma^l_k(i))P(\vec{\sigma})\right].\ee
Injecting Eq.~(\ref{APP:FINIE:eq:def:pNM1H1...}) into the latter equation, we find
\be \label{APP:FINIE:eq:P:C:inter1}\begin{split} d_tP_{N_G}(\vec{\Sigma})=\sum_{l=1, 2}\sum_{k=1}^{N_G}\sum_{\vec{\sigma}\in \mathcal{S}(\vec{\Sigma})}\sum_{i=0}^{n_b-1}\sum_{\sigma=\pm1}w_l(k, i, \sigma)\delta(\sigma+\sigma^l_k(i))P(\vec{\sigma}^l_k(i))\\
-\sum_{l=1, 2}\sum_{k=1}^{N_G}\sum_{\vec{\sigma}\in \mathcal{S}(\vec{\Sigma})}\sum_{i=0}^{n_b-1}\sum_{\sigma=\pm1}w_l(k, i, \sigma)\delta(\sigma-\sigma^l_k(i))P(\vec{\sigma}).\end{split}\ee
Spins that are in box $k$ but do not contribute to $M_k$ because they are outside the vicinity of the central spin give a zero contribution due to the sum on $\vec{\sigma}$. We can therefore transform $\sum_{i=1}^{n_{b}-1}$ into $\sum_{i=1}^{n(a)}$ where we recall that we labeled from $i=0$ to $i=n(a)$ the spins (and fields) belonging to the neighborhood of the central spin (and field) $i=0$.

\begin{figure}
    \centering
    \includegraphics{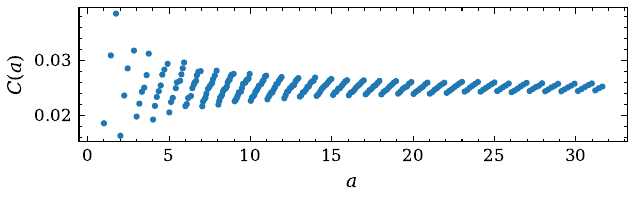}
    \caption{Parameter $C$ introduced in Eq.~(\ref{APP:FINIE:eq:C}) as a function of the interaction range $a$ for a cubic lattice.}
    \label{fig:ca}
\end{figure}

We introduce
\be A_k^l(\vec{\Sigma})=\sum_{i=0}^{n(a)}\sum_{\vec{\sigma}\in \mathcal{S}(\vec{\Sigma})}\delta(\sigma^l_k(i)-\sigma)w_l(k,i,\sigma)P(\vec{\sigma})\ee
corresponding to the last term of the previous equation without the sum over $l$, $k$ and $\sigma$. 
We seek to write the first term of Eq.~(\ref{APP:FINIE:eq:P:C:inter1}) as a function of $A_k^l$.
We denote $\Delta_k^l$ the operator calculating the difference when flipping a spin ($l=1$) or a field ($l=2$) $\sigma$ in box $k$. We have $\Delta_k^l \vec{\Sigma}=-\sigma\vec{d}_k^l/n(a)$ with 
\begin{align}\vec{d}^1_{k}=&(0, \dots,\underbrace{2}_{k\text{th position}}, \dots, 0)\in \mathbb{R}^{2N_G}, \vec{d}^2_{k}=&(0, \dots,\underbrace{2}_{ \text{position }N_G+k}, \dots, 0)\in \mathbb{R}^{2N_G}.\end{align}
We transform the sum on the configurations $\vec{\sigma}\in \mathcal{S}(\vec{\Sigma})$ with $\sigma_k^l(i)=-\sigma$ into a sum on configurations $\vec{\sigma}'\in \mathcal{S}(\vec{\Sigma}+\sigma \vec{d}_k^l/n(a))$ with a spin (or field) $\sigma_k^l(i)=+\sigma$ of a configuration $\vec{\sigma}'$.
Eq.~(\ref{APP:FINIE:eq:P:C:inter1}) thus becomes 
\be d_t P_{N_G}(\vec{\Sigma})=\sum_{l=1, 2}\sum_{k=1}^{N_G}\sum_{\sigma=\pm1}
\left[A_k^l\left(\vec{\Sigma}+\frac{\sigma \vec{d}_{l, k}}{n(a)}\right)-A_k^l(\vec{\Sigma})\right].\ee
We now consider a limit of large $n(a)$, which allows us to do a perturbative expansion in $1/n(a)$ of $A_k^l\left(\vec{\Sigma}+\frac{\sigma \vec{d}_{l, k}}{n(a)}\right)$.
We keep only the terms up to order $1/n(a)$, leading to the following equation:
\be \label{APP:FINIE:eq:Ei:Fi}\begin{split} d_t P_{N_G}(\vec{\Sigma})=\sum_{k=1}^{N_G}\left[\underset{E_1}{\underbrace{\sum_{\sigma=\pm1}
\frac{2\sigma}{n(a)}\partial_{M_k}A_k^1(\vec{\Sigma})}}+\underset{F_1}{\underbrace{\sum_{\sigma=\pm1}\frac{2}{n(a)^2}\partial^2_{M_k}A_k^1(\vec{\Sigma})}}\right.\\
+\left.\underset{E_2}{\underbrace{\sum_{\sigma=\pm1}\frac{2\sigma}{n(a)}\partial_{H_k}A_k^2(\vec{\Sigma})}}+\underset{F_2}{\underbrace{\sum_{\sigma=\pm1}\frac{2}{n(a)^2}\partial^2_{H_k}A_k^2(\vec{\Sigma})}}\right].\end{split}\ee
In the previous equation, we introduced four different terms $E_1, E_2, F_1$ and $F_2$, which we now seek to simplify.
We assume that $M_k(i)\approx M_k$ and $H_k(i)\approx H_k$ and keep only the first order in $M_k(i)-M_k$ and in $H_k(i)-H_k$. 
We thus have 
\be w_1(k, i, \sigma)\approx \frac{1}{1+e^{2\sigma (J_1M_k+ H_k)/T}}-\frac{2\sigma e^{2\sigma (J_1M_k+H_k)/T } [J_1(M_k(i)-M_k)+H_k(i)-H_k]}{T(1+e^{2\sigma (J_1M_k+H_k)/T})^2}.\ee
We first consider the lowest order term in $M_k(i)-M_k$ and $H_k(i)-H_k$. 
This term corresponds to the transition rates for a mean-field model inside the box $k$, since the transition rates no longer depend on the position $i$ in the box. Its contribution to the $E_1$ term is therefore 
\be  [M_k-\tanh(H_k/T)] P_{N_G}(\vec{\Sigma}).\ee
For the higher-order term in $M_k(i)-M_k$ and $H_k(i)-H_k$, it is assumed that $M_k$ and $H_k$ are small, which is valid near the transition temperature only, such that  
\be \frac{2\sigma e^{2\sigma(J_1M_k+ H_k)/T } }{T(1+e^{2\sigma(J_1M_k+ H_k)/T})^2}\sim \frac{\sigma}{2T}.\ee
The contribution of the higher-order term in $M_k(i)-M_k$ and $H_k(i)-H_k$ to the $E_1$ term thus becomes
\be-\frac{1}{Tn(a)}\sum_{\sigma=\pm1}\sum_{i=1}^{n(a)} \sum_{\vec{\sigma}\in \mathcal{S}(\vec{\Sigma})}\delta(\sigma^1_k(i)-\sigma)\sigma^2\left[J_1(M_k(i)-M_k)+H_k(i)-H_k\right]P(\vec{\sigma})\ee 
which, summed over $\sigma=\pm1$, becomes 
\be-\frac{1}{Tn(a)}\sum_{i=1}^{n(a)} \sum_{\vec{\sigma}\in \mathcal{S}(\vec{\Sigma})}\left[J_1(M_k(i)-M_k)+H_k(i)-H_k\right]P(\vec{\sigma}).\ee 
We wish to write $M_k(i)-M_k$ and $H_k(i)-H_k$ solely as a function of $M_k$ and $H_k$.
We introduce a position variable $\mathbf{x}$ such that two boxes are separated by a distance $|\Delta \mathbf{x}|=1$. Two neighboring spins are then separated by the distance $|\mathbf{x}(i)-\mathbf{x}(j)|=1/(2\lfloor a\rfloor+1)$. We write $\mathbf{x}_k$ the position of the central spin ($i=0$) of box $k$. We write $\mathbf{x}_k+\delta \mathbf{x}_k(i)$ the position of spin $i$ in box $k$. 
We have :
\be M_k(i)-M_k
=\!\!\!\sum_{m\in\{1, 2,3\}}\!\!\! \delta\mathbf{x}_k(i)\cdot \mathbf{u}_m \partial_m M(\mathbf{x}_k)+\frac{1}{2}\!\!\sum_{m, n\in\{1, 2, 3\}}\!\!\!\!\delta\mathbf{x}_k(i)\cdot \mathbf{u}_m\partial_m\partial_nM(\mathbf{x}_k)\delta\mathbf{x}_k(i)\cdot\mathbf{u}_n,\ee
where $\mathbf{u}_m$ is a unit vector oriented along the $m\in\{1,2,3\}$ axis in $3$-dimensional space. For the linear and quadratic terms corresponding to $m\neq n$, the sum over $i$ is zero by symmetry. So we have :
\be\frac{1}{n(a)}\sum_{i=1}^{n(a)} (M_k(i)-M_k)
=\sum_{m\in\{1, 2, 3\}}\frac{1}{2n(a)}\sum_{i=1}^{n(a)}(\delta\mathbf{x}_k(i)\cdot \mathbf{u}_m)^2\partial_m^2 M(\mathbf{x}_k),\ee
and we introduce 
\be \label{APP:FINIE:eq:C} C= \frac{1}{2n(a)}\sum_{i=1}^{n(a)}(\delta \mathbf{x}_k(i)\cdot \mathbf{u}_1)^2\ee 
which, by symmetry, leads to the following equation
\be\frac{1}{n(a)}\sum_{i=1}^{n(a)} (M_k(i)-M_k)
=C\nabla^2 M(\mathbf{x}_k).\ee
Finally, we get 
\be E_1=\big[M_k-\tanh\left[(J_1M_k+H_k)/T\right]- C\Delta (J_1M_k+H_k)/T \big] P_{N_G}(\vec{\Sigma})\ee
and similarly, we can simplify $E_2$ into
\be\begin{split} E_2=\left[H_k-\tanh[(J_2H_k+(1-\mu)M_k)/T]- C\Delta (J_2H_k+(1-\mu)M_k)/T\right] P_{N_G}(\vec{\Sigma}).\end{split}\ee
We come back to the quantity $C$ introduced in equation~(\ref{APP:FINIE:eq:C}), which depends on the lattice geometry. For a cubic lattice, it depends only on the range of the interactions $a$, see Fig.~\ref{fig:ca}. 
We find that $C$ depends little on $a$ and $C\approx 0.025\pm 0.005$. As explained in the main text, we thus neglect the dependence of $C$ on $a$ and take $C=0.025$. 

To simplify $F_1$ and $F_2$, we consider the lowest order in $(M_k, H_k)$, leading to 
\be F_1=F_2=\frac{1}{n(a)}\ee
which simplifies Eq.~(\ref{APP:FINIE:eq:Ei:Fi}) into the following equation 
\be \begin{split}\partial_t  &P_{N_G}(\vec{\Sigma}=(M_1,...,M_{N_G} H_1, ..., H_{N_G}))=\\
\sum_{k=1}^{N_G} \Bigg[&\partial_{M_k}\left(M_k-\tanh\left[\frac{J_1M_k+H_k}{T}\right]-\frac{C}{T}\Delta (J_1M_k+H_k)\right) P_{N_G}(\vec{\Sigma})\\
+&\partial_{H_k}\left(H_k-\tanh\left[\frac{J_2H_k+(1-\mu)M_k}{T}\right]-\frac{C}{T}\Delta (J_2H_k+(1-\mu)M_k)\right) P_{N_G}(\vec{\Sigma})\\
+&\frac{1}{n(a)}\partial_{M_k}^2 P_{N_G}(\vec{\Sigma})+\frac{1}{n(a)}\partial_{H_k}^2P(\vec{\Sigma})\Bigg].\end{split}\ee
This equation corresponds to the Fokker-Planck equation on $P_{N_G}(\vec{\Sigma})$ given in Eq.~(\ref{eq:FP:res}).

\subsection{Mean-field coupling of noisy oscillators}
\label{app:mean:field:complement}
In this appendix, we detail how the distribution $P_1(r, \theta)$ of local variables in a given box, with mean-field couplings, was obtained in Eq.~(\ref{eq:p1rtheta}).
We start from the three-dimensional noisy equation on $z_k$ [Eq.~(\ref{eq:zk:final})] :
\be\frac{dz_k}{dt}=(\gamma-1)z_k-|z_k|^2z_k+\gamma C\Delta z_k+\sqrt{D}\left[\eta_k^R+i\eta_k^I\right]. \ee
Replacing the Laplacian by a mean-field coupling [see Eq.~(\ref{eq:MF:approx})], and introducing $z_k=x_k+iy_k$, leads to
 \be \begin{split}\frac{dx_k}{dt}&=(\gamma-1)x_k-(x_k^2+y_k^2)x_k+\gamma c\left(\overline{x}-x_k\right)+\sqrt{D}\eta_k^R\\
\frac{dy_k}{dt}&=(\gamma-1)y_k-(x_k^2+y_k^2)y_k+\gamma c\left(\overline{y}-y_k\right)+\sqrt{D}\eta_k^I.\end{split}\ee
 These equations are Langevin equations associated with the following Fokker-Planck equation:
 \be \label{APP:FINIE:eq:FP:xy}\begin{split}&\frac{\partial P_{N_G}(\vec{x}, \vec{y})}{\partial t}\\&=-\sum_k \partial_{x_k}[(\gamma-1)x_k-(x_k^2+y_k^2)x_k+\gamma c\left(\overline{x}-x_k\right)]P_{N_G}(\vec{x}, \vec{y})\\
&-\sum_k \partial_{y_k}[(\gamma-1)y_k-(x_k^2+y_k^2)y_k+\gamma c \left(\overline{y}-y_k\right)]P_{N_G}(\vec{x}, \vec{y})\\
&+\frac{D}{2}\sum_k\left[\partial_{x_k}^2+\partial_{y_k}^2\right]P_{N_G}(\vec{x}, \vec{y})\end{split}\ee
with the notations $\vec{x}=(x_1, ..., x_{N_G})$ and $\vec{y}=(y_1, ..., y_{N_G})$.
We take the limit $N_G\to\infty$, and since we are considering the limit where all boxes interact, we introduce $P_1(x, y)$, the distribution of variables in a single box, such that
\be \label{eq:app:factorization}
P_{N_G}(\vec{x}, \vec{y})=\prod_{k=1}^{N_G} P_1(x_k, y_k).
\ee
By injecting the ansatz (\ref{eq:app:factorization}) into Eq.~(\ref{APP:FINIE:eq:FP:xy}), we obtain an equation on $P_1(x, y)$:
\be\begin{split}\frac{\partial P_1(x, y)}{\partial t}=& -\partial_{x}\left[(\gamma-1)x-(x^2+y^2)x+\gamma c\int dx' dy' P_1(x', y')(x'-x)\right]P_1(x, y)\\
& -\partial_{y}\left[(\gamma-1)y-(x^2+y^2)y+\gamma c \int dx' dy' P_1(x', y')(y'-y)\right]P_1(x, y)\\
&+\frac{D}{2}\left[\partial_{x}^2+\partial_{y}^2\right]P_1(x, y).\end{split}\ee
The stationary solution of this equation is:
\be \label{APP:FINIE:eq:p1xy}\begin{split}P_1(x, y)=p_0 \exp\left[-\frac{2}{D}\left(\frac{1}{4}(x^2+y^2)^2-\frac{\gamma-1}{2}(x^2+y^2)+\frac{\gamma c}{2}\left((\overline{x}-x)^2+(\overline{y}-y)^2\right)\!\right)\!\right]\end{split}\ee
where, by definition of $\overline{z}$, 
\be\label{APP:FINIE:eq:self:consistent} \overline{z}=\overline{x}+i\overline{y}=\int dxdy (x+iy) P_1(x, y)\ee
which corresponds to a self-consistent equation.  
In polar coordinates, we write $\overline{z}=Re^{i\Theta}$, $z=r e^{i\theta}$, which leads to Eqs.~(\ref{eq:p1rtheta}) and (\ref{eq:self:consistent:2}) of the main text.

\section{Comment on the influence of $J_1$}
\label{app:J1}
\begin{figure}[t]
    \centering
\includegraphics{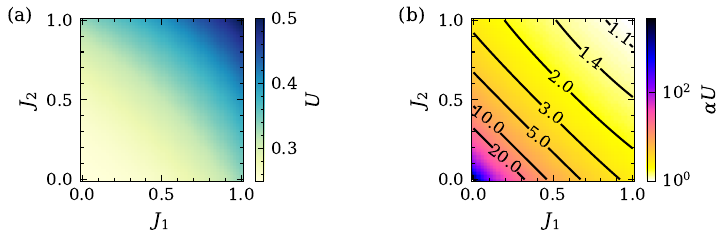}
    \caption{Influence of $J_1$ and $J_2$ on parameters (a) $U$ controlling oscillation amplitude and (b) $\alpha U$ controlling noise amplitude, for $\mu=2$. }
    \label{APP:FINIE:fig:D:J1etJ2}
\end{figure}
Throughout this paper, we have considered $J_1=0$ in order to reduce the space of parameters to be studied. 
On Fig.~\ref{APP:FINIE:fig:D:J1etJ2}, we plot in color the amplitude of $U$ and $\alpha U$ in the $(J_1,J_2)$-plane for $\mu=2$. For a given value of $J_2$, increasing $J_1$ means increasing $\alpha U$ and therefore lowering the transition temperature. As $U$ increases slightly, the amplitude of the magnetization slightly decreases, when all boxes are synchronized. We can see that the coupling constant $J_1$ plays a similar role to the coupling constant $J_2$, so that considering $J_1=0$ is not a strong restriction.

\bigskip

\bibliographystyle{plain_url}
\bibliography{biblio}

\end{document}